\documentclass[twocolumn,showpacs,preprintnumbers,amsmath,amssymb]{revtex4}

\usepackage{graphicx}
\usepackage{dcolumn}
\usepackage{lscape}
\usepackage{bm}
\usepackage{color}

\begin{document}
\preprint{APS/123-QED}

\title{Itinerant ferromagnetism in actinide $5f$ electrons system: \\Phenomenological analysis with spin fluctuation theory\footnote{Phys. Rev. B. {\bf 96}, 035125 (2017).}}

\author{Naoyuki Tateiwa$^{1}$}
\email{tateiwa.naoyuki@jaea.go.jp} 
\author{Ji\v{r}{\'i} Posp{\'i}\v{s}il$^{1,2}$}
\author{Yoshinori Haga$^{1}$}%
\author{Hironori Sakai$^{1}$}%
\author{Tatsuma D. Matsuda$^{3}$}%
\author{Etsuji Yamamoto$^{1}$}

\affiliation{
$^{1}$Advanced Science Research Center, Japan Atomic Energy Agency, Tokai, Naka, Ibaraki 319-1195, Japan\\
$^2$Charles University in Prague, Faculty of Mathematics and Physics, Department of Condensed Matter Physics, Ke Karlovu 5, 121 16 Prague 2, Czechia\\
$^3$Department of Physics, Tokyo Metropolitan University, Hachioji, Tokyo 192-0397, Japan\\
}
\date{\today}

\begin{abstract}
We have carried out an analysis of magnetic data in 69 uranium, 7 neptunium and 4 plutonium ferromagnets with the spin fluctuation theory developed by Takahashi (Y. Takahashi, J. Phys. Soc. Jpn. {\bf 55}, 3553 (1986)). The basic and spin fluctuation parameters of the actinide ferromagnets are determined and the applicability of the spin fluctuation theory to actinide $5f$ system has been discussed. Itinerant ferromagnets of the $3d$ transition metals and their intermetallics follow a generalized Rhodes-Wohlfarth relation between ${p_{\rm eff}}/{p_{\rm s}}$ and ${T_{\rm C}}/{T_0}$, viz., ${p_{\rm eff}}/{p_{\rm s}}{\,}{\propto}{\,}({T_{\rm C}}/{T_0})^{-3/2}$. Here,  $p_{\rm s}$, $p_{\rm eff}$, $T_{\rm C}$, and $T_0$ are the spontaneous and effective magnetic moments, the Curie temperature and the width of spin fluctuation spectrum in energy space, respectively. The same relation is satisfied for ${T_{\rm C}}/{T_0}< 1.0$ in the actinide ferromagnets. However, the relation is not satisfied in a few ferromagnets with ${T_{\rm C}}/{T_0}{\,}{\sim}{\,}1.0$ that corresponds to local moment system in the spin fluctuation theory. The deviation from the theoretical relation may be due to several other effects not included in the spin fluctuation theory such as the crystalline electric field effect on the $5f$ electrons from ligand atoms. The value of the spontaneous magnetic moment $p_{\rm s}$ increases linearly as a function of ${T_{\rm C}}/{T_0}$ in the uranium and neptunium ferromagnets below $(T_{\rm C}/{T_0})_{\rm {kink}}{\,}{=}{\,}0.32{\,}{\pm}{\,}0.02$ where a kink structure appears in relation between the two quantities.  $p_{\rm s}$ increases more weakly above $(T_{\rm C}/{T_0})_{\rm {kink}}$. A possible interpretation with the ${T_{\rm C}}/{T_0}$-dependence of $p_{\rm s}$ is given. 
 \end{abstract}


\maketitle

\section{Introduction}
Actinide compounds with $5f$ electrons have long attracted much attention because of their interesting magnetic and electronic properties such as heavy fermion features, unconventional superconductivity, co-existence of the superconductivity and magnetism, and physical phenomena associated with multipole degrees of freedom of the the $5f$ electrons\cite{santini1,santini2,moore,stewart}. Similar unusual physical properties have been extensively studied in other strongly correlated electrons systems such as oxides and organic and rare earth compounds. It is necessary to reveal the behavior of the electrons responsible for these properties and find universality of the electronic properties in the different systems.

The peculiarities of the physical phenomena in actinide systems are ascribed to the role of the $5f$ electrons with larger spatial extent than that of the $4f$ electrons in the lanthanides. The electrons show a tendency to delocalization through strong hybridization between $5f$ orbitals and conduction states. In actinide metallic compounds, the degree of localization of the $5f$ electrons differs in different compounds, ranging from strongly localized to itinerant character. The $3d$ electrons in transition metals also show various degree of localization. Differences between the $5f$ and $3d$ electrons are the smaller sensitivity to the crystal field (CF) from ligand atoms and the stronger spin-orbit coupling in the $5f$s. The interplay of the spin and the unquenched orbital degrees of freedom gives rise to the peculiar features of the observed physical phenomena in actinide compounds. 

While many theoretical studies have been done for the role of the $5f$ electrons on their interesting physical properties\cite{santini1,santini2,moore}, the behavior of the $5f$ electrons has not been fully elucidated yet. When the $5f$ electrons have strongly localized character at higher temperature, a Kondo-lattice picture may be appropriate for understanding the formation of the strongly correlated electric states at low temperatures as has been established in the $4f$ electrons system of the rare-earth cerium (Ce) and ytterbium (Yb) compounds\cite{thalmeier1,hewson,fazekas}. Certainly, some actinide compounds show behaviors reminiscent of the Kondo effect such as the logarithmic temperature dependence of the electrical resistivity (${\rho}{\,}{\sim}${\,}-ln$T$). But not all physical properties in actinide compounds have been consistently explained on this point of view. There has been no experimental report of continuous and systematic evolution from the Kondo impurity to the concentrated Kondo lattice system, in contrast to the rare earth Ce system such as CeAl$_2$\cite{onuki1}, CeB$_6$\cite{nksato0}, and CeCu$_6$\cite{onuki2} whose physical properties can be continuously tuned by replacing the Ce ion with La without $4f$ electron. Furthermore, there have been controversies as to whether the $5f$ electrons should be treated as being itinerant or localized in various uranium compounds. The readers refer to Refs. 1 and 11 for these issues in the uranium chalogenide compounds (USe, US and UTe) and UAsSe\cite{santini1,vogt1}.  Generally, neither theoretical models based on the assumption of the localized nor itinerant $5f$ electrons can explain the physics of actinide metallic compounds. From a different point of view, this duality in the nature of the $5f$ electrons has been positively taken as a starting point in theoretical models for the heavy fermion superconductors UPd$_2$Al$_3$ or UPt$_3$\cite{thalmeier1,sato,zwicknagl,efremov}. 

 In this study, we focus on the nature of the $5f$ electrons in actinide ferromagnets and present interesting views deduced from analysis of the magnetic data using spin fluctuation theory. Among the actinide ferromagnets, uranium ferromagnetic superconductors UGe$_2$, URhGe and UCoGe have been extensively studied both theoretically and experimentally for more than 10 years with the interest arising from the  same uranium $5f$ electrons underlying both long-range ordered states\cite{saxena,huxley0,aoki0,huy,aoki2,aoki3,huxley1}. Ferromagnetic superconductivity has been an important research subject since its theoretical prediction by Ginzburg in 1956\cite{ginzburg}. The study of the $5f$ electrons in the uranium ferromagnetic superconductors has potential importance not only for actinide science but also for wide research fields of the superconductivity and the magnetism.
 
 In the theoretical studies for the ferromagnetic superconductors, the ferromagnetism of the $5f$ electrons has been variously approached via mean field theory\cite{sandeman}, the spin-fermion model\cite{karchev1,karchev2}, the Hubbard, periodic Anderson or Kondo Hamiltonians\cite{kaneyasu,kubo1,wysokinski,coqblin1,coqblin2}, and the $j$-$j$ coupling scheme\cite{hirohashi,hotta1}. Some of the physical properties of the ferromagnetic superconductors have been explained by these theories but there have remained unsolved problems. One source of the difficulties can be attributed to the diversity of the nature in the $5f$ electrons as mentioned above. Photoemission spectroscopy suggests $5f$ electrons' itinerant character in UGe$_2$, URhGe and UCoGe\cite{fujimori}. But the Ising-type anisotropy in the ferromagnetic state favors a localized model of the $5f$s. The degree of the itinerancy of the $5f$s may differ between the ferromagnetic superconductors. The dualism of $5f$s in UGe$_2$ has been discussed in both experimental and theoretical studies\cite{karchev1,karchev2,yaouanc,sakarya,troc0}. The complexity of the $5f$s makes it difficult to find an appropriate starting point for the electronic state of these electrons in theoretical studies, making it necessary to find a method to evaluate the degree of the itinerancy of the of the $5f$s in actinide ferromagnets. 
    
  In this paper, we report results of a phenomenological analysis on actinide ferromagnets using the spin fluctuation theory that was developed to account for the ferromagnetic properties of itinerant ferromagnets in the $3d$ transition metals and their intermetallics\cite{moriya1}. In the spin fluctuation theory, the degree of the itinerancy of the magnetic electrons is defined by the parameter ${T_{\rm C}}/{T_0}$, where $T_0$ is the width of spin fluctuation spectrum in energy space. We adapt this theory and find a theoretical relation for the basic magnetic properties that also holds for actinide $5f$ electrons ferromagnets with certain exceptions. The parameter ${T_{\rm C}}/{T_0}$ can also here be used to estimate the degree of the itinerancy of the $5f$s. We also suggest that relevant parameters for the actinide ferromagnetism are related to the itinerancy of the $5f$s, although some properties characteristic of a localized nature may play a role in the magnetic properties. 
     
  This paper is organized as follow. In the Sec. II, we briefly summarize the history of the spin fluctuation theory and the theoretical framework for the analysis of magnetic data in actinide ferromagnets. The experimental methods and analyses are given in Sec. III. In Secs. IV and V, we present the results of our analysis on 80 actinide ferromagnets and a summarizing discussion, respectively. A summary is given in Sec. VI.

\section{Spin fluctuation theory}
\subsection{Brief history}
The theoretical study of itinerant electron ferromagnets can be traced back to works by Bloch\cite{bloch}, Slater\cite{slater} and Stoner\cite{stoner} starting in the late 1920s. Slater discussed the ferromagnetism of nickel (Ni) with tight binding $d$ bands with intra-atomic exchange interaction only and Stoner developed the itinerant electron theory of ferromagnetism at finite temperature on the basis of the Hartree-Fock and mean-field approximations, respectively. The early theories explain ground-state magnetic properties of itinerant ferromagnets such as not integer values of the observed spontaneous magnetic moment in the $3d$ metals and their intermetallics. But there were difficulties in explaining finite-temperature magnetic properties such as Curie-Weiss behavior of the magnetic susceptibility above the ferromagnetic transition temperature $T_{\rm C}$. These come from mean-field treatment of the on-site Coulomb interaction $U$ whose strength ($\sim$ 10 eV) is one or two orders of magnitude larger than that of the kinetic energy $K$ of the conduction electrons. Theoretical studies have been done to overcome the difficulties. 

We focus on the improved mean-field approximation which takes into account the spatial spin density fluctuations. The collective nature of magnetic excitations was considered in Herring and Kittel's theory of the spin waves by using the random phase approximation (RPA)\cite{herring1,herring2}. Izuyama, Kim, and Kubo developed the RPA theory for the dynamical magnetic susceptibility taking into account dispersive collective excitations: the spin fluctuations. The theory shows the magnetic critical scattering around $T_{\rm C}$ and the spin-wave dispersion in the ferromagnetic state\cite{izuyama}. The RPA theory was developed into the paramagnon model for nearly ferromagnetic metals or liquid helium-3 ($^3$He)\cite{berk,doniach1,bealmond}. These RPA theories are effective at low temperatures. Moriya and Kawabara developed self-consistent renormalization (SCR) theory where the dynamical susceptibility and the free energy are renormalized in a self-consistent way\cite{moriya1,moriya2,moriya3}. Phenomenological mode-mode coupling theory, equivalent to the SCR theory at high temperatures, was proposed by Murata and Doniach\cite{murata}. These advances were followed by alternative derivations of the SCR theory\cite{ramakrishnan,hertz2,kuroda,dzyaloshinskii,lonzarich} and then extended to cover antiferromagnetic metals\cite{hasegawa0}. There are two important results found: One is a substantial reduction of $T_{\rm C}$ from the value in the Stoner's mean field theory, and the other a new mechanism for the Curie-Weiss magnetic susceptibility in a wide temperature region. Other results from spin fluctuation theory have been confirmed in a number of experimental studies such as inelastic neutron scattering experiment on the weak itinerant ferromagnet MnSi\cite{ishikawa1}.

In reality, most of metallic ferromagnets are located in an intermediate region between the local moment and weakly coupling limits that the early theories assumed. A unified theory interpolating the two limiting cases has been developed by Moriya and Takahashi using a functional integral method\cite{moriya5}. Quantitative calculations of the ferromagnetic properties with electronic band structures were performed using the functional integral method\cite{hasegawa1,hasegawa2}. First-principles approaches to the electronic structure of magnetic materials have also advanced remarkably for several decades and important progress has been made in understanding finite-temperature magnetic properties of the itinerant ferromagnets iron (Fe) and Ni with the combination of the dynamic mean-field theory (DMFT) and electric structure calculations\cite{kotliar1,kotliar2,kotliar3,kotliar4}.  

From the late 1980's, the spin fluctuation theory has been extended to study the unconventional superconductivity with anisotropic superconducting order parameter in strongly correlated electron systems such as high-$T_c$ cuprates, heavy fermion superconductors, and two-dimensional organic compounds\cite{moriya6,pines1}. A theoretically predicted linear relation between the superconducting transition temperature $T_{\rm {sc}}$ and the spread of the spin fluctuation spectrum in energy space $T_0$ has been confirmed in the cuprate and heavy fermion superconductors\cite{moriya7,nksato1}. Moriya's SCR theory was derived using of renormalization group theory\cite{millis} and re-formulated with the standard $s$-$f$ model to explain the anomalous physical properties around magnetic instability in the heavy fermion systems of the rare earth and actinide compounds\cite{moriya8}. The validity of the theory has been confirmed by careful experimental studies\cite{kambe1,kambe2,kambe3} and is regarded as one of standard views in the research field of quantum critical phenomena\cite{lohneyseh}. In this study, we apply the spin fluctuation theory to actinide ferromagnets.

\subsection{Takahashi's spin fluctuation theory}
 We analyze the magnetic data of uranium (U), neptunium (Np), and plutonium (Pu) ferromagnets with the spin fluctuation theory developed by Takahashi\cite{takahashi1,takahashi2,takahashi3,takahashi4}. The total amplitude of the local spin fluctuation ${\langle}{S_{\rm L}^2}{\rangle}_{\rm {total}}$ is composed of the thermal ${\langle}{S_{\rm L}^2}{\rangle}_{\rm {T}}$ and the zero point fluctuations ${\langle}{S_{\rm L}^2}{\rangle}_{\rm {Z. P.}}$. The Takahashi's spin fluctuation theory assumes that ${\langle}{S_{\rm L}^2}{\rangle}_{\rm {total}}$ is constant as a function of temperatures. This assumption is reasonable, considering the weak temperature dependence of the spin amplitude in the order of $T_{\rm C}$ seen in neutron-scattering experiments on MnSi\cite{ziebeck} and Y$_{0.97}$Sc$_{0.03}$Mn$_2$\cite{shiga}, and theoretical calculations with the Hubbard models\cite{shiba1,shiba2,hirsch,nakano} and the Moriya's SCR theory\cite{moriyatakahashi}. The effectiveness of the theory has been confirmed in a number of experimental studies on intermetallic compounds of the $3d$ transition metals\cite{yoshimura1,yoshimura2,nishihara,shimizu,nakabayashi,chen1,ohta,yang1,yang2,chen2,imai,waki1}. 
  
 The local spin fluctuation density squared ${\langle}{S_{\rm L}^2}{\rangle}$ is related to the imaginary part of the dynamical magnetic susceptibility ${\chi}({\mbox{\boldmath $q$}},{\omega})$ through the fluctuation and dissipation theorem. The spin fluctuation spectrum for itinerant ferromagnets is described by double Lorentzian distribution functions in small energy $\omega$ and wave vector {\mbox{\boldmath $q$}} spaces\cite{takahashi3}. 
   \begin{eqnarray}
{\rm Im} {\chi}({\mbox{\boldmath $q$}},{\omega}) &=& {{{\chi}(0)}\over 1+q^2/{\kappa}^2}{{\omega}{{\Gamma}_{q}}\over {\omega}^2+{\Gamma}_{q}^2}\\
{{\Gamma}_{q}}{\,}&=&{\,}{{\Gamma}_{0}}q({\kappa}^2+q^2)
 \end{eqnarray}
Here, $q{\,}{\equiv}{\,}{\lvert}{\mbox{\boldmath $q$}}{\rvert}$, ${\kappa}$ represents the inverse of the magnetic correlation length, and ${{\Gamma}_{q}}$ is the damping constant, the inverse of the life time of the fluctuation with wave vector {\mbox{\boldmath $q$}}. The spectrum is represented in a parameterized form by introducing two energy scales $T_0$ ($={{\Gamma}_0}q_{\rm B}^3/{2{\pi}}$) and $T_{\rm A}$ ${[={N_0}{q_{\rm B}^2}/(2{\chi}(0){{\kappa}^2}})]$, where $q_{\rm B}$ is the zone-boundary wave vector for the crystal with $N_0$ magnetic atoms with the volume $V$ ($={6{\pi}^2}{N_0}/{q_{\rm B}^3}$). These parameters represent the distribution widths of the spin fluctuation spectrum in the energy and wave-vector spaces, respectively.  

In the Takahashi's spin fluctuation theory, the Landau expansion of free energy $F{_m}(M)$ is expressed as 
     \begin{eqnarray}
 F{_m}(M)&=&F{_m}(0)+ {1\over 2}a(0)M^2+{1\over 4}b(0)M^4  \\
a(0)&=& {{1}\over {{(g{\mu_{\rm B}})^2{{\chi}(0)}}}} \nonumber \\
b(0)&=&{{F_1}\over{(g{\mu_{\rm B}})^4{N_0^3}}} \nonumber
  \end{eqnarray} 
where $g$ represents Lande's $g$ factor. $F_1$ is the mode-mode coupling term, the coefficient of the $M^4$ term. 

The magnetization $M$ at the ground state is expressed by the following equation
  \begin{eqnarray}
H &=& {{F_1}\over {{N_0^3}{(g{\mu_{\rm B}})^4}}}(-{M_0^2}+M^2)M\\
F_1&=& {{{2}{{T_{\rm A}}^2}}\over {15cT_0}}
 \end{eqnarray}   
where $c$ = 1/2, and $M_0$ is the spontaneous magnetic moment. The parameter $F_1$ is connected with $T_0$ and $T_{\rm A}$ through Eq. (5) and can be evaluated experimentally from the inverse slope of the Arrott plots ($M^2$ versus $H/M$ plot) at low temperatures, 
  \begin{equation} F_1={{{N_{\rm A}}^3}(2{{\mu}_{\rm B}})^4\over {k_{\rm B}}{\zeta}}\end{equation}
where $N_{\rm A}$ is Avogadoro's number and $k_{\rm B}$ is the Boltzmann constant, and ${\zeta}$ is the slope of the Arrott plot ($H/M$-$M^2$ curve) at low temperatures\cite{arrott}. $T_0$ and $T_{\rm A}$ are expressed with $F_1$ in following relations: 
   \begin{eqnarray}
 &&{\left({{T_{\rm C}}\over{T_0}}\right)^{5/6}} = {{p_{\rm s}^2}\over {5{g^2}C_{4/3}}} {\left({15c{F_1}\over{{2}{T_{\rm C}}}}\right)^{1/2}}\\
&&  {\left({{T_{\rm C}}\over{T_{\rm A}}}\right)^{5/3}} = {{p_{\rm s}^2}\over {5{g^2}C_{4/3}}} {\left({{2}{T_{\rm C}}\over{15c{F_1}}}\right)^{1/2}}
 \end{eqnarray}
 where $C_{4/3}$ is a constant ($C_{4/3}$ = 1.006089${\,}{\cdot}{\cdot}{\cdot}$). $p_{\rm s}$ is the spontaneous magnetic moment and $T_{\rm C}$ is the ferromagnetic transition temperature expressed in following equations:
  \begin{eqnarray}
 p_{\rm s}^2 &=& {{12{T_0}}\over {T_{\rm A}}}{C_{4/3}} {\left({{T_{\rm C}}\over{T_0}}\right)^{4/3}}\\
 {T_{\rm C}}&=& (60c)^{-3/4}{p_{\rm s}^{3/2}}{{T}_{\rm {A}}^{3/4}}{{T}_{0}^{1/4}}
 \end{eqnarray}

 An important consequence from the Takahashi's theory is that the parameters $F_1$, $T_0$, and $T_{\rm A}$ can be determined from experimental magnetic data only. Meanwhile, the parameters are independent in the Moriya's SCR theory and the neutron or nuclear magnetic resonance spectroscopy is necessary to determine them\cite{moriya1,moriya2,moriya3}. 
 
A ratio ${T_{\rm C}}/{T_0}$ characterizes the degree of itinerancy of magnetic electrons in the spin fluctuation theory. At ${T_{\rm C}}/{T_0}$ $\ll$ 1, the magnetic electrons have a strong itinerant character. The value of ${p_{\rm eff}}/p_{\rm s}$ is large and ${T_{\rm C}}/{T_0}$ becomes small for weak ferromagnets with large spin fluctuation amplitude. Both quantities approach to unity when the degree of itinerancy of the magnetic electrons becomes small. The local magnetic moment is responsible for the ferromagnetism when ${T_{\rm C}}/{T_0}{\,}={\,}1$.

 In 1963, Rhodes and Wohlfarth proposed to plot the ratio ${p_{\rm eff}}/{p_{\rm s}}$ as a function of $T_{\rm C}$ (Rhodes and Wohlfarth plot)\cite{rhodes, wohlfarth}. The ratio ${p_{\rm eff}}/{p_{\rm s}}$ was defined as a measure of quantification to the degree of itinerancy of the magnetic electrons. This Rhodes and Wohlfarth plot has been widely used in studies of itinerant ferromagnets for a long time. However, there is no theoretical ground to the relation between ${p_{\rm eff}}/{p_{\rm s}}$ and $T_{\rm C}$. In the Takahashi's theory, ${p_{\rm eff}}/{p_{\rm s}}$ is described as a function of the parameter ${T_{\rm C}}/{T_0}$ as follows:
 
   \begin{eqnarray}
{{p_{\rm eff}}\over{{p_{\rm s}}}}{\,}&=&{\,}{\left({{1}\over {10{C_{4/3}}{\rm d}y/{\rm d}t}}\right)^{-1/2}}{\left({{T_{\rm C}}\over{{T_{0}}}}\right)^{-2/3}} \nonumber \\ 
 &{\simeq}&{\,}1.4{\left({{T_{\rm C}}\over{{T_{0}}}}\right)^{-2/3}}
   \end{eqnarray}
 where $y$ is the inverse magnetic susceptibility and $t$ is the reduced temperature $t$ = ${T}/{T_0}$. The quantity ${[1/({10{C_{4/3}}{\rm d}y/{\rm d}t})]}^{-1/2}$ is numerically estimated as $\sim$ 1.4\cite{takahashi3}. This generalized Rhodes-Wohlfarth relation between ${p_{\rm eff}}/{{p_{\rm s}}}$ and ${T_{\rm C}}/{T_0}$ has been experimentally confirmed in a number of ferromagnetic compounds in the $3d$ electrons systems\cite{yoshimura1,yoshimura2,nishihara,shimizu,nakabayashi,chen1,ohta,yang1,yang2,chen2,imai,waki1}. The spin fluctuation parameters for the uranium feromagnetic superconductors UGe$_2$, URhGe and UCoGe have been reported by Deguchi, Takahashi, and Sato\cite{takahashi3,nksato2}. In this paper, we discuss the applicability of the spin fluctuation theory to the actinide $5f$ electrons system from the analyses on the magnetic data of 80 actinide ferromagnets with the Takahashi's theory.

\section{Methods of experiment and analysis}
  We have examined the 69 uranium, 7 neptunium, and 4 plutonium ferromagnets listed in Tables I, II, and III. The uranium ferromagnets are divided into two categories, group I and group II. We have obtained the basic magnetic and the spin fluctuation parameters from the analyses of our experimental data for the uranium ferromagnets in group I. The single crystal samples of group I were grown by the Czochralski crystal pulling method. The magnetic measurements have been done with a commercial superconducting quantum interference device (SQUID) magnetometer. The parameters for the uranium ferromagnets in group II, and neptunium and plutonium ferromagnets are obtained from the analyses of experimental data taken from the literature.
  \begin{figure}[t]
\includegraphics[width=8.5cm]{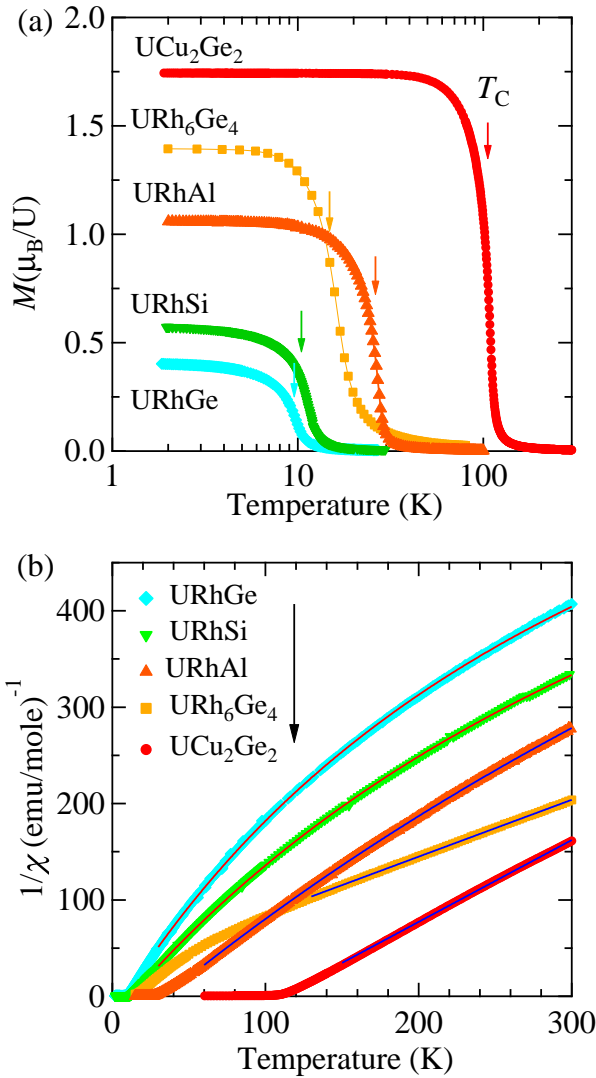}
\caption{\label{fig:epsart}Temperature dependencies of (a) the magnetization $M$ under magnetic field and (b) the inverse of the magnetic susceptibility $1/{\chi}$ for several uranium ferromagnets UCu$_2$Ge$_2$, URhAl, URh$_6$Ge$_4$, URhSi, and URhGe in magnetic fields of 0.5, 0.2, 0.1, 0.1 and 0.1 T, respectively, applied along the magnetic easy axes. The Curie temperatures $T_{\rm C}$ are denoted by arrows. Solid lines are results of fits to the temperature dependence of the inverse of the magnetic susceptibility $1/{\chi}$ with the Curie-Weiss or modified Curie-Weiss law.}
\end{figure}

 \begin{figure}[]
\includegraphics[width=8.5cm]{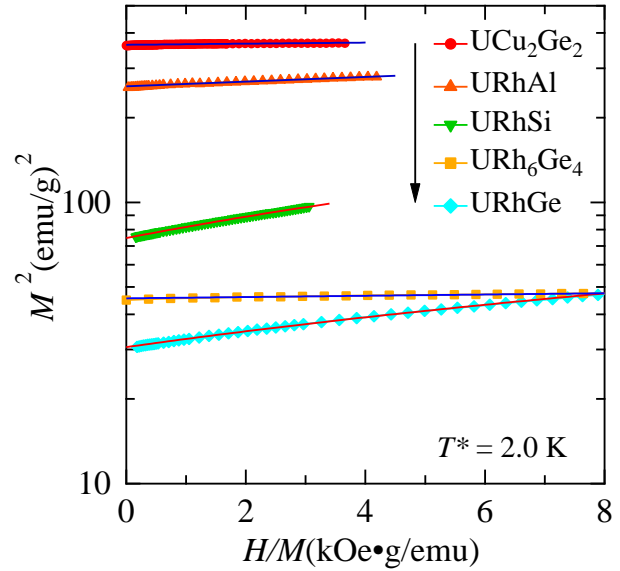}
\caption{\label{fig:epsart}$M^2$ versus $H/M$ plot (Arrott plots) at $T^*$ = 2.0 K for UCu$_2$Ge$_2$, URhAl, URh$_6$Ge$_4$, URhSi, and URhGe in applied magnetic field along the magnetic easy axes.}
\end{figure}  
  We selected the ferromagnetic compounds that show a single ferromagnetic phase transition and excluded ferromagnets like AnFe$_2$ (An: U, Np, and Pu) where the magnetic moment of the $3d$ transition metal has an important contribution in the magnetic property at zero magnetic field\cite{wulff}. Spin fluctuation theory assumes a simple ferromagnetic state. We excluded ferromagnets with complex magnetic structures such as U$_3$P$_4$ and U$_3$As$_4$\cite{santini1}. Generally, the ferromagnetic state has magnetic anisotropy in actinide compounds. We have analyzed the magnetic data of the actinide ferromagnets obtained using single crystal samples in applied magnetic field along the magnetic easy axis. There are some exceptions as will be mentioned in the next section.

   The parameter $F_1$ is determined from the slope ${\zeta}$ of the Arrott-plot ($H/M$-$M^2$ curve) at $T^*$ with the Eq. (6). Then the spin fluctuation parameters $T_0$ and $T_{\rm A}$ can be estimated with the value of $p_{\rm s}$ using Eqs. (7) and (8). We confirm that the transition temperature $T_{\rm C}$ determined experimentally is reproduced from the values of $p_{\rm s}$, $T_0$ and $T_{\rm A}$ using Eq. (10). The effective magnetic moment $p_{\rm eff}$ is determined from the slope of the inverse magnetic susceptibility $1/{\chi}$ from the Curie-Weiss (CW) law  ${\chi}$ = ${C/(T-{\theta})}$. Here, $C$ is the Curie constant and $\theta$ the paramagnetic Curie temperature. The effective magnetic moment, $p_{\rm eff}$, per magnetic atom is estimated from $C$ = ${N_0}{{\mu}_{\rm B}^2}{p_{\rm eff}^2}/3{k_{\rm B}}$. In some ferromagnets, $1/{\chi}$ is not linear in temperature. The magnetic susceptibility $\chi$ was then analyzed with the modified Curie-Weiss (mCW) law, ${\chi}$ = ${C/(T-{\theta})}+ {{\chi}_0}$. $ {{\chi}_0}$ is a temperature-independent term which may arise from the density of states at the Fermi energy other than $5f$ electrons.

\section{Results}

\begin{table*}[t]
\caption{\label{tab:table1}%
Basic magnetic and spin fluctuation parameters for uranium ferromagnets in group I. These parameters are obtained with present authors' experimental data taken on single crystalline samples and unpublished data are used for the ferromagnets marked with an asterisk *. The spontaneous magnetic moment $p_{\rm s}$ and the effective magnetic moment ${p_{\rm eff}}$ are estimated from the magnetization at $T^*$ and the magnetic susceptibility, respectively. The spin fluctuation parameters, $F_1$, $T_0$, and $T_{\rm A}$ are estimated with Takahashi's spin fluctuation theory. The mode-mode coupling term $F_1$ is determined from the magnetization data at $T^*$. Abbreviation CW or mCW in column ${\chi}(T)$ denotes that the magnetic susceptibility has been analyzed the Curie-Weiss or modified Curie-Weiss law, respectively.}
\begin{ruledtabular}
\begin{tabular}{ccccc|cccc|cc|cc}
\textrm{}&
\textrm{$T_{\rm C}$}&
\textrm{$p_{{\rm eff}}$ }&
\textrm{$p_{\rm_s}$}&
\textrm{${p_{\rm eff}}/p_{\rm s}$}&
\textrm{$F_1$}&
\textrm{$T_0$}&
\textrm{$T_{\rm A}$}&
\textrm{${T_{\rm C}}/{T_0}$}&
\textrm{${\chi}(T)$}&
\textrm{$T^*$}&
\textrm{Ref.}&\\
\textrm{}&
\textrm{(K)}&
\textrm{(${\mu}_{\rm B}/{\rm U}$)}&
\textrm{(${\mu}_{\rm B}/{\rm U}$)}&
\textrm{}&
\textrm{(K)}&
\textrm{(K)}&
\textrm{(K)}&
\textrm{}&
\textrm{}&
\textrm{(K)}&
\textrm{}&\\
\colrule
\multicolumn{3}{l}{Uranium compounds: group I}&&&&&&&&&\\
UGe$_2$&52.6    &3.00   & 1.41  & 2.13 & 554 & 92.2 &442 &0.571&CW&2.0& \cite{galatanu1,tateiwa1}  \\
UIr&46.0  & 3.40&0.492  &6.91&1.97$\times$10$^3$&440&1.80$\times$10$^3$ &0.105&CW&2.0& \cite{galatanu2}\\
UGa$_2$&123  &3.30   &2.94 &1.12&273&94.8&311&1.12&CW&2.0&\cite{honma}\\
URhGe$_2$ & 30.0  & 3.06&0.768&3.99& 517  &170&574&0.176&CW&2.0&\cite{matsuda1}  \\
UCu$_2$Ge$_2$  &109   &2.93&1.74&1.69&521&187&605&0.582&CW&2.0&\cite{matsuda2}\\
URhGe  &9.47    &1.75  & 0.407  & 4.30 &1.10$\times$10$^3$& 78.4 & 568 &0.121 &mCW &2.0&\cite{tateiwa1} \\
URh$^*$&57&2.26&0.652&3.47&1.52$\times$10$^3$&367&1.45$\times$10$^3$&0.155&mCW&2.0&\\
URh$_6$Ge$_4$$^*$&14.8  & 3.58&1.39  &  2.58 &560&13.2&167&1.15&CW&2.0&\\
URhAl$^*$&26.2&2.50&1.05&2.37&428&71.8&340&0.365&mCW&2.0&\\
URhSi$^*$ &10.5    &2.94 & 0.571  & 5.15  &520 &64.5&354&0.163&CW&2.0& \\
URh$_{1-x}$Co$_x$Ge$^*$ &&&&&&&&&&&\\
$x=0.2$ &13.7   & 1.86  &  0.450 &4.13&1.22$\times$10$^3$ &104&691&0.131&mCW&1.87&\\
$x=0.6$ &19.7   & 1.91  &  0.498 &3.83&1.01$\times$10$^3$ &164&788&0.120&mCW&1.87&\\
$x=0.7$&18.6&1.94&0.416&4.65&945&239&921&0.0777&mCW&1.87&\\
$x=0.8$ &15.0   &  1.92 &  0.293 &6.56&1.11$\times$10$^3$&358&1.22$\times$10$^3$&0.0419&mCW&1.87&\\
$x=0.9$ &7.0   &  1.94  & 0.127 &15.3&2.93$\times$10$^3$&439&2.19$\times$10$^3$&0.0160&mCW&1.87&\\
URh$_{1-x}$Ir$_x$Ge$^*$ &&&&&&&&&&&\cite{jirka2}\\
$x=0.15$ &9.3   &  1.75 &  0.392 &4.47&1.22$\times$10$^3$&78.3&599&0.119&mCW&1.87&\\
$x=0.43$ &6.0   &  1.73  & 0.292 &5.92&1.27$\times$10$^3$&76.6&605&0.0783&mCW&1.87&\\
\end{tabular}
\end{ruledtabular}
 \end{table*}

 \begin{table*}[]
\caption{\label{tab:table1}%
Basic magnetic and spin fluctuation parameters for uranium ferromagnets in group II. These parameters are estimated from the analysis of experimental data from the literatures. The meanings of the other notations are the same as in Table I.}
\begin{ruledtabular}
\begin{tabular}{ccccc|cccc|cc|cc}
\textrm{}&
\textrm{$T_{\rm C}$}&
\textrm{$p_{{\rm eff}}$ }&
\textrm{$p_{\rm_s}$}&
\textrm{${p_{\rm eff}}/p_{\rm s}$}&
\textrm{$F_1$}&
\textrm{$T_0$}&
\textrm{$T_{\rm A}$}&
\textrm{${T_{\rm C}}/{T_0}$}&
\textrm{${\chi}(T)$}&
\textrm{$T^*$}&
\textrm{Ref.}&\\
\textrm{}&
\textrm{(K)}&
\textrm{(${\mu}_{\rm B}$/U)}&
\textrm{(${\mu}_{\rm B}$/U)}&
\textrm{}&
\textrm{(K)}&
\textrm{(K)}&
\textrm{(K)}&
\textrm{}&
\textrm{}&
\textrm{(K)}&
\textrm{}&\\
\hline
\multicolumn{3}{l}{Uranium compounds: group II}&&&&&&&&&\\
UPt&28.6   & 3.58& 0.822&4.36&566&127&519&0.226&CW&5.0& \cite{prokes}\\
US &177&2.35&1.51&1.56&1.71 $\times$ 10$^3$&275&1.34 $\times$ 10$^3$&0.643&mCW&4.05&\cite{tillwick,bourdarot}\\
USe &160&2.50&1.72&1.45&823&271&914&0.591&mCW&4.2&\cite{busch1}\\
UTe &104&2.70&1.90&1.42&537&139&528&0.751&mCW&1.5&\cite{busch2}\\
U$_5$Sb$_4$&86&2.98&1.63&1.83&416&171&517&0.503&CW&1.5&\cite{paixao} \\
UAsS&124&3.34&1.17&2.85&1.08 $\times$ 10$^3$&383&1.25 $\times$ 10$^3$&0.323&CW&4.2&\cite{zygmunt,bazan} \\
UAsSe&113&3.41&1.29&2.64&957&281&1.00 $\times$ 10$^3$&0.403&CW&4.2&\cite{zygmunt,bazan} \\
UAsTe&66&3.34&1.29&2.59&343&222& 534 &0.298&CW&4.2&\cite{zygmunt,bazan} \\
UPS&118&2.57&1.04&2.47&1.16 $\times$ 10$^3$&454&1.40 $\times$ 10$^3$&0.260&CW&5.0&\cite{kaczorowski3}\\
UPSe&55&3.17&1.03&3.06&363&271&607&0.203&CW&5.0&\cite{kaczorowski4}\\
UPTe&85&2.83&1.37&2.06&421&252&631&0.337&CW&5.0&\cite{kaczorowski4}\\
USbSe&128&3.08&1.68&1.83&905&189&802&0.676&CW&5.0&\cite{kaczorowski4}\\
USbTe&127&3.18&1.92&1.65&1.11 $\times$ 10$^3$&120&705&1.06&CW&5.0&\cite{kaczorowski4}\\
USeTe&69&3.04&1.59&1.91&197&201&385&0.343&mCW&5.0&\cite{troc1}\\
USTe&85&2.92&1.54&1.89&325&223&522&0.381&mCW&5.0&\cite{troc1}\\
UCu$_{0.9}$Sb$_2$ &113  &3.10  & 1.67 &1.88&414&252&93.5&0.448&CW&1.9&\cite{bukowski1} \\
UCo$_{0.5}$Sb$_2$ &65  &2.80  & 1.11 &2.52&429&271&660&0.240&CW&1.9&\cite{bukowski2} \\
UAuBi$_{2}$ &22.5  &3.30 & 1.25&2.64&223&55.3&215&0.407&CW&1.8&\cite{rosa} \\
UCuAs$_2$&133&2.68&1.27&2.11&5.29 $\times$ 10$^3$&136&1.64 $\times$ 10$^3$&0.979&mCW&4.2&\cite{kaczorowski1} \\
UCuP$_2$&74.5&2.21&0.993&2.23&1.09 $\times$ 10$^4$&63.3&1.61$\times$ 10$^3$&1.18&mCW&4.2&\cite{kaczorowski1} \\
UCu$_2$P$_2$ &216&2.26&1.80&1.25&2.15 $\times$ 10$^3$&219&1.33 $\times$ 10$^3$&0.987&mCW&4.2&\cite{kaczorowski2}\\
U$_3$TiSb$_5$&160&2.81&1.65&1.70&3.07 $\times$ 10$^3$&135&1.25 $\times$ 10$^3$&1.18&CW&5.0&\cite{mar}\\
U$_3$ScSb$_5$&130&2.86&1.46&1.95&688&318&906&0.409&CW&5.0&\cite{mar}\\
U$_3$Cu$_4$Ge$_4$ &73.0   & 2.99&1.68  &1.77&183&200&370&0.365&mCW&2.0&\cite{gorbunov}\\
U$_3$Fe$_4$Ge$_4$ &18.0   & 1.93&0.402   &4.80&1.27$\times$10$^3$&207&993&0.0871&mCW&2.0&\cite{henriques1}\\
U$_2$Fe$_3$Ge &55   & 2.52&0.488   &5.16&4.54$\times$10$^3$&362&2.48$\times$10$^3$&0.152&mCW&2.0&\cite{henriques2} \\
U$_2$RhSi$_3$&24.0  &2.50  & 0.707&3.53&635&128&552&0.187&mCW&1.72&\cite{szlawska}\\
U$_3$Co$_2$Ge$_7$ &40&2.41&1.07&2.25&900&87.2&543&0.459&mCW&2.0&\cite{uhlirova} \\
U$_4$(Ru$_{1-x}$Os$_x$)$_7$Ge$_6$  &&&&&&&&&&&\cite{colineau}\\
$x=0$ &12.0    &1.38&  0.206 &6.71&8.79$\times$10$^3$&170&2.36$\times$10$^3$&0.0708&mCW&2.0&\\
$x=0.1$ &9.0   &1.36  & 0.148&9.20&8.20$\times$10$^3$&246&2.75$\times$10$^3$&0.0365&mCW&2.0&\\
$x=0.2$ &7.0   &1.35  & 0.118 &11.4&7.88$\times$10$^3$&289&2.92$\times$10$^3$&0.0312&mCW&2.0&\\
UCoGa&47  &2.40 & 0.638&3.76&1.44 $\times$ 10$^3$&295&1.26 $\times$ 10$^3$&0.159&CW&4.2&\cite{nakotte1,purwanto} \\
UPtAl&43.5  &2.85 & 1.38&2.06&615&67.8&395&0.642&CW&4.2&\cite{andreev2}  \\
UIrAl &62  &2.20 & 0.960&2.30&820&241&861&0.257&CW&4.2&\cite{andreev3} \\
UCoAl$_{0.75}$Sn$_{0.25}$ &5.5  &2.66 & 0.169&15.7&924&300&1.02$\times$10$^3$&0.0183&mCW&2.0&\cite{andreev1}\\
UCo$_{1-x}$Ru$_x$Al&&&&&&&&&&&\cite{valivska1,jirka1} \\
$x=0.005$&4.5   &  1.87  & 0.322 &5.81&1.24$\times$10$^3$&39.0&426&0.115&mCW&1.8& \\
$x=0.01$ &16   &  1.83  & 0.365 &5.02&2.19$\times$10$^3$&156&1.13$\times$10$^3$&0.103&mCW&1.8&\\
$x=0.62$ &38.0   &  1.90  & 0.428 &4.44 &2.54$\times$10$^3$&396&1.94$\times$10$^3$&0.0972&mCW&1.8&\\
$x=0.70$&17.0   &  2.20  & 0.169 &13.0 &3.93$\times$10$^3$&766&3.36$\times$10$^3$&0.0222&mCW&1.8& \\
$x=0.74$&6.5   &  2.00  & 0.0732 &27.3 &8.78$\times$10$^3$&759&5.00$\times$10$^3$&0.00856&mCW&1.8& \\
UCo$_{1-x}$Os$_x$Al &&&&&&&&&&&\cite{andreev4}\\
$x=0.02$ &26   &  1.91  & 0.394 &4.85&1.74$\times$10$^3$&324&1.45$\times$10$^3$&0.0802&mCW&2.0&\\
$x=0.1$&48   &  1.95  & 0.518 &3.766&1.84$\times$10$^3$&433&1.73$\times$10$^3$&0.111&mCW&2.0& \\
\end{tabular}
\end{ruledtabular}
 \end{table*}

      \begin{table*}[]
\caption{\label{tab:table1}%
Basic magnetic and spin fluctuation parameters for uranium ferromagnets in group II, neptunium and plutonium ferromagnets. These parameters are estimated from the analysis of experimental data from the literatures. The parameters are estimated with data taken on polycrystalline samples for the ferromagnets marked with a dagger $\dagger$.  An indicates actinide atom U, Np, or Pu. $^{\star}$The value of $p_{\rm eff}$ is assumed to be 1.90 in UCo$_{1-x}$Fe$_x$Ge (see text). $^{\ddagger}$The parameters in UCoGe are cited from Refs. \cite{takahashi3,nksato2}. The meanings of the other notations are the same as in Table I. }
\begin{ruledtabular}
\begin{tabular}{ccccc|cccc|cc|cc}
\textrm{}&
\textrm{$T_{\rm C}$}&
\textrm{$p_{{\rm eff}}$ }&
\textrm{$p_{\rm_s}$}&
\textrm{${p_{\rm eff}}/p_{\rm s}$}&
\textrm{$F_1$}&
\textrm{$T_0$}&
\textrm{$T_{\rm A}$}&
\textrm{${T_{\rm C}}/{T_0}$}&
\textrm{${\chi}(T)$}&
\textrm{$T^*$}&
\textrm{Ref.}&\\
\textrm{}&
\textrm{(K)}&
\textrm{(${\mu}_{\rm B}$/An)}&
\textrm{(${\mu}_{\rm B}$/An)}&
\textrm{}&
\textrm{(K)}&
\textrm{(K)}&
\textrm{(K)}&
\textrm{}&
\textrm{}&
\textrm{(K)}&
\textrm{}&\\
\colrule
\multicolumn{3}{l}{Uranium compounds: group II}&&&&&&&&&\\
UCo$_{1-x}$Ru$_x$Ge &&&&&&&&&&&\cite{valivska1,valivska2}\\
$x=0.03$ &6.50   &  1.88  & 0.108 &17.4&7.96$\times$10$^3$&316&3.07$\times$10$^3$&0.0206&mCW&1.8&\\
$x=0.10^{\dagger}$ &8.62   & 1.86  &  0.111 &16.8&1.38$\times$10$^4$ &337&4.18$\times$10$^3$&0.0256&mCW&1.8&\\
$x=0.12$&7.46   &  1.88 &  0.149 &12.6&5.54$\times$10$^3$&226&1.44$\times$10$^3$&0.0330&mCW&1.8& \\
$x=0.23^{\dagger}$ &4.68   & 1.83  &  0.0544 &34.6&2.43$\times$10$^4$&495&6.72$\times$10$^3$&0.00945&mCW&1.8&\\
UCo$_{1-x}$Fe$_x$Ge$^{{\dagger},{\star}}$ &&&&&&&&&&&\cite{huang1}\\
$x=0.025$ &7.73   & 1.9  & 0.156 &12&3.99$\times$10$^3$&251&1.94$\times$10$^3$&0.0308&mCW&2.0&\\
$x=0.05$ &8.35   & 1.9  & 0.206 &9.2&2.00$\times$10$^3$&229&1.31$\times$10$^3$&0.0364&mCW&2.0&\\
$x=0.075$ &8.55   & 1.9  & 0.310 &6.1&606&183&645&0.0466&mCW&2.0&\\
$x=0.10$ &6.90   & 1.9  & 0.229 &8.3&1.07$\times$10$^3$&157&875&0.0389&mCW&2.0&\\
$x=0.125$ &7.02   & 1.9  & 0.220 &8.8&763&263&869&0.0267&mCW&2.0&\\
$x=0.15$ &6.40   & 1.9  & 0.0980 &19&3.25$\times$10$^3$&805&2.85$\times$10$^3$&0.00894&mCW&2.0&\\
UCoGe$^{\ddagger}$&2.4&1.93&0.039&49.5&2.87$\times$ 10$^4$&362&5.92$\times$ 10$^3$&0.0065&mCW&0.1&\cite{takahashi3,nksato2}\\
\colrule
\multicolumn{3}{l}{Neptunium compounds}&&&&&&&&&\\
NpAl$_2$&56.0  &2.40&1.21  & 1.99 &884&113&613&0.494&CW&4.2&\cite{aldred0,aldred1}\\
NpOs$_2$&7.50   & 3.10 &0.361 & 8.58 &589&105&481&0.0715&CW&4.2&\cite{aldred0,aldred1}  \\ 
NpNi$_2$&32.0   &2.42  &0.756& 3.19  &519&195&616&0.164&CW&5.0&\cite{aldred2} \\
Np$_2$C$_3^{\dagger}$ &109   &2.40&0.920 & 2.61&1.53$\times$10$^3$&452&1.61$\times$10$^3$&0.241&CW&4.2&\cite{lam}\\
NpSb$_2$&47.0   &2.52  & 1.57&1.60&282&90.0&309&0.522&CW&5.0&\cite{homma}  \\
NpNiSi$_2$&51.5   &2.16  & 1.13&1.92&848&120&617&0.427&CW&5.0&\cite{colineau2}  \\
NpFe$_4$P$_{12}$&23   &1.55  & 1.35&1.15&1.33$\times$10$^3$&16.4&285&1.40&mCW&5.0&\cite{aoki1}  \\
\colrule
\multicolumn{3}{l}{Plutonium compounds}&&&&&&&&&\\
PuAs&125  &1.00&0.655& 1.53 &2.85$\times$10$^4$ &232&4.98$\times$10$^3$&0.557&CW&4.2&\cite{mattenberger}\\
PuP &126  &1.06  & 0.618&1.72&6.04$\times$10$^4$&164&6.09$\times$10$^3$&0.770&mCW&4.2&\cite{vogt1,lam2}\\
PuGa$_3^{\dagger}$ ($R$-3$m$)&20  &0.77 &0.197 &3.91&5.92$\times$10$^4$&136&5.49$\times$10$^3$&0.147&mCW&5.0&\cite{boulet}\\
Pu$_2$Pt$_3$Si$_5$&58.0  &0.78 &0.305 &2.56&6.95$\times$10$^4$&249&7.79$\times$10$^3$&0.242&mCW&5.0&\cite{griveau}\\
\end{tabular}
\end{ruledtabular}
 \end{table*}
       \begin{figure*}[t]
\includegraphics[width=14cm]{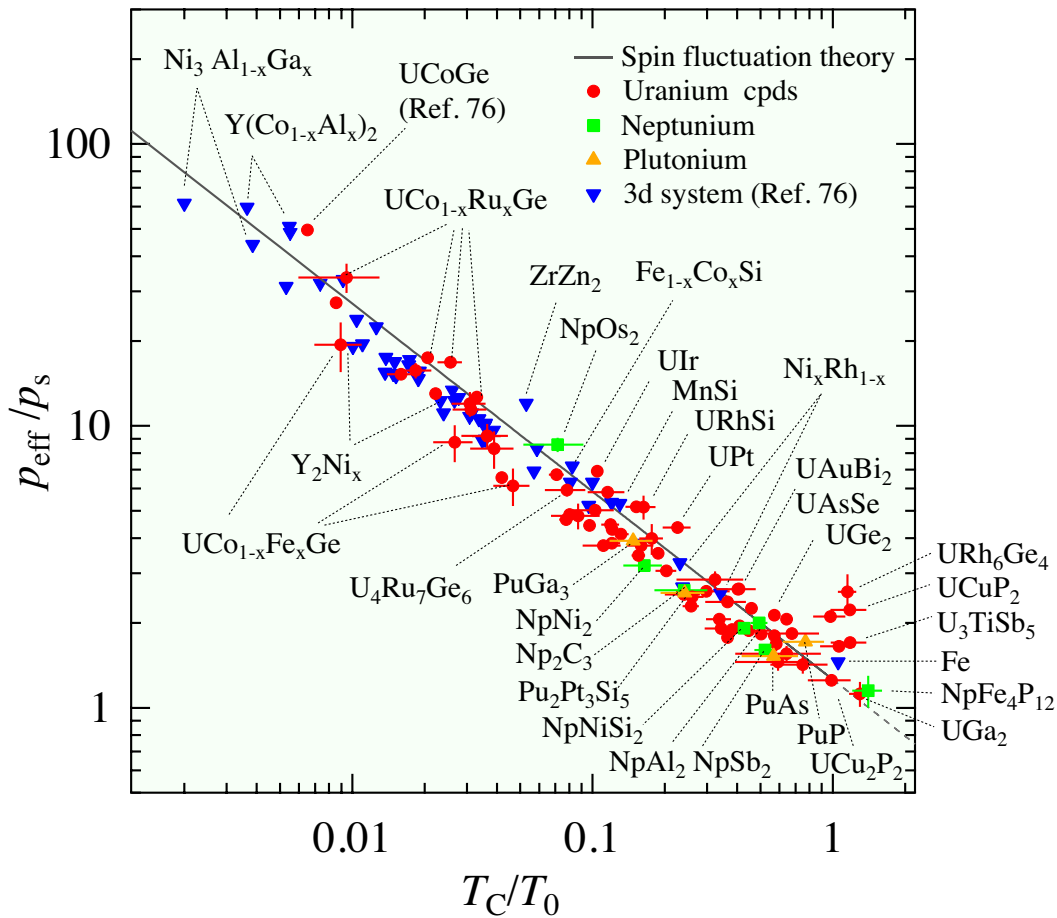}
\caption{\label{fig:epsart}(Color online)Generalized Rhodes-Wohlfarth plot. Data points for uranium, neptunium and plutonium compounds are plotted as closed circles, squares and triangles, respectively. The data of UCoGe are cited from Ref. 76. Data for intermetallic ferromagnetic compounds of the transition $3d$ metals cited from Refs\cite{takahashi1,takahashi2,takahashi3} are represented as closed antitriangles. The solid line is the theoretical relation between ${T_{\rm C}}/{T_0}$ and ${p_{\rm eff}}/p_{\rm s}$, Eq. (11) in Takahashi's spin fluctuation theory\cite{takahashi3}.}
\end{figure*} 
       \begin{figure}[]
\includegraphics[width=8.5cm]{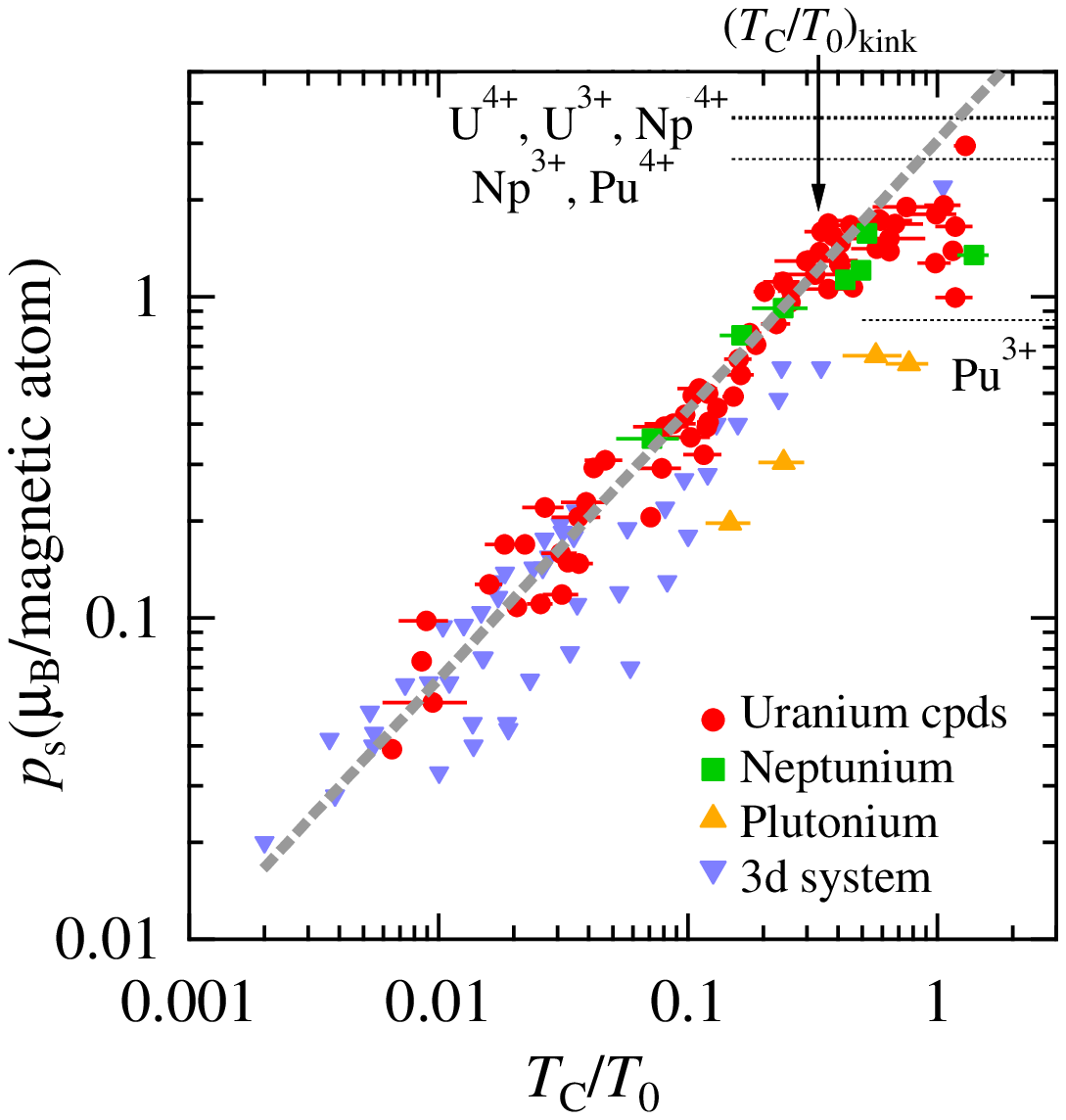}
\caption{\label{fig:epsart}Relations between the spontaneous magnetic moment $p_{\rm s}$ and ${T_{\rm C}}/{T_0}$ for uranium, neptunium, and plutonium compounds plotted as closed circles, squares, and triangles, respectively. Data points for intermetallic ferromagnetic compounds of the transition $3d$ metals represented as antitriangles are cited from Refs. 41, 42, and 43. The dotted line is a fit to the data of the uranium and neptunium ferromagnets for $T_{\rm C}/{T_0}{\,}{<}{\,}(T_{\rm C}/{T_0})_{\rm {kink}}{\,}{=}{\,}0.32{\,}{\pm}{\,}0.02$ with the function $p({T_{\rm C}}/{T_0})=a({T_{\rm C}}/{T_0})^{-n}$ where $a$ and $n$ are fitting parameters.}
\end{figure} 

\subsection{Basic magnetic and spin fluctuation parameters and generalized Rhodes-Wohlfarth plot}
  Table I shows the basic magnetic and spin fluctuation parameters of the uranium ferromagnets in group I. These have been determined from the analysis of our experimental data. We used published data of UGe$_2$\cite{galatanu1,tateiwa1}, URhGe\cite{tateiwa1}, UIr\cite{galatanu2}, UGa$_2$\cite{honma}, URhGe$_2$\cite{matsuda1}, UCu$_2$Ge$_2$\cite{matsuda2}, and URh$_{1-x}$Ir$_x$Ge\cite{jirka2}. Unpublished data were analyzed for the ferromagnets URh, URh$_6$Ge$_4$, URhSi, URhAl, and URh$_{1-x}$Co$_x$Ge marked with an asterisk * in the table. The detailed studies of the physical properties in the latter compounds are now in progress using high-quality single crystal samples and part of the magnetic data were used for this study. Among them, URh$_6$Ge$_4$ is a new ferromagnet with ${T_{\rm C}}$ = 14.8 K that crystalizes in the hexagonal LiCo$_6$P$_4$-type structure. The ferromagnetic property in URh, isostructural to UIr, has been previously studied using polycrystal samples\cite{onuki3,nishioka}. We have studied the anisotropic ferromagnetic properties in URh using single crystal samples. URhSi and URhAl are known ferromagnets\cite{veenhuizen,honda,prokes2}. The basic magnetic properties of the ferromagnets such as the value of $T_{\rm C}$ or $p_{\rm s}$ in our single crystal samples are consistent with those in previous studies. We are interested in doping effect on the uranium ferromagnetic superconductors URhGe and UCoGe. The systematic study of the magnetic properties in the series of URh$_{1-x}$Ir$_x$Ge and URh$_{1-x}$Co$_x$Ge are underway with high-quality single crystal samples. The doping dependencies of the spin fluctuations parameters are shown in this paper. Note that the basic magnetic properties in the series are consistent with those in previous studies using polycrystal samples\cite{chevalier,huy2}. 
  
 Figure 1 (a) shows the temperature dependencies of the magnetization $M$ and (b) the inverse of the magnetic susceptibility $1/{\chi}$ for several uranium ferromagnets UCu$_2$Ge$_2$, URhAl, URh$_6$Ge$_4$, URhSi, and URhGe in the group I in magnetic fields of 0.5, 0.2, 0.1, 0.1, and 0.1 T, respectively, applied along the magnetic easy axes of the ferromagnets. The Curie temperature $T_{\rm C}$ denoted by an arrow is determined from the temperature dependences of the specific heat, the electrical resistivity, and the magnetization in zero or low magnetic field. The effective magnetic moment $p_{\rm eff}$ is determined from the fit to the magnetic susceptibility ${\chi}$ with the Curie-Weiss or modified Curie-Weiss law as shown in lines in Fig. 1(b). The obtained values of the parameters, $T_{\rm C}$, $p_{\rm eff}$, and $p_{\rm s}$ are listed in Table I. Generally, the values of $p_s$ and $p_{\rm eff}$ are smaller than those expected for $5f^2$ (U$^{4+}$, $p_{\rm eff}$ = 3.58 ${\mu}_{\rm B}$/U) and $5f^3$ (U$^{3+}$, $p_{\rm eff}$ = 3.62 ${\mu}_{\rm B}$/U) configurations, which can be understood as manifestation of the itinerancy of the $5f$ electrons. 
 
 Figure 2 shows the $M^2$ versus $H/M$ plot (Arrott plots) at $T^*$ = 2.0 K for UCu$_2$Ge$_2$, URhAl, URh$_6$Ge$_4$, URhSi, and URhGe in applied magnetic field along the magnetic easy axes of the ferromagnets. We estimate $F_1$ from the slope ${\zeta}$ of Arrot plots shown as lines in the figure. The obtained values of $F_1$ are 521, 428, 560, 520, and 1.10 $\times$ 10$^3$ K, respectively. Then, the spin fluctuation parameters $T_0$ and $T_{\rm A}$ are estimated with Eqs. (7) and (8) using the values of $T_{\rm C}$, $F_1$, and $p_{\rm s}$. These obtained parameters are also listed in the Table I. We have estimated the parameters for the rest of the uranium ferromagnets in the group II in the same ways and show them in Table II. The spin fluctuation parameters for UGe$_2$ and URhGe in this study are consistent with those determined previously\cite{takahashi3,nksato2}. 
  
 Extensive experimental studies have been carried out on many actinide ferromagnets more than five decades. To examine applicability of the spin fluctuation theory to the actinide system from a wider viewpoint, we decided to analyze the magnetic data of actinide ferromagnets reported in literature. We selected ferromagnets according to the criteria mentioned in the previous section and analyzed the magnetic data from litereture. The basic magnetic and the spin fluctuation parameters are obtained for 52 uranium ferromagnets, 7 neptunium and 4 plutonium ferromagnets as listed in Table II and III. Note that the parameters in UCoGe determined previously are cited from Refs. \cite{takahashi3,nksato2}.  As can be seen from the Table I, II, and III,  $T_{\rm C}/{T_0}$ in the actinide ferromagnets shows a wide range of values from $T_{\rm C}/{T_0}$ = 0.0065 for UCoGe to 1.70 for U$_3$TiSb$_3$. The degree of itinerancy of $5f$ electrons in the actinide ferromagnets largely differ, depending on each ferromagnet. It is interesting to compare the values of $T_{\rm C}/{T_0}$ in the actinide ferromagnets with those of well-known itinerant ferromagnets in the $3d$ system such as Ni$_3$Al ($T_{\rm C}/{T_0}$ = 0.015), MnSi (0.13), ZrZn$_2$ (0.053),  Y(Co$_{1-x}$Al$_x$)$_2$ (0.0036${\,}{\sim}{\,}$0.018), Ni (0.237), and Fe (1.05)\cite{takahashi1,takahashi3}.
 
 Figure 3 shows the double logarithmic plot of ${p_{\rm eff}}/p_{\rm s}$ and ${T_{\rm C}/T_0}$ (the generalized Rhodes-Wohlfarth plot) for all actinide ferromagnets analyzed. The data points for the uranium, neptunium, and plutonium ferromagnets are shown in closed circles, squares, and triangles, respectively. Solid line is the theoretical relation [Eq. (11): ${p_{\rm eff}/p_{\rm s}}{\,}={\,}1.4({T_{\rm C}}/T_0)^{-2/3}$] in the Takahashi's spin fluctuation theory. For comparison, we plot the data of the $3d$ transition metals and their intermetallics such as Ni$_3$Al,  Ni$_3$Al$_{1-x}$Ga$_x$ (0 $\le$ $x$ $\le$ 0.33), (Fe$_{1-x}$Co$_x$)$_3$Mo$_3$N, Sc$_3$In, Y(Co$_{1-x}$Al$_x$)Co$_2$ (0.13 $\le$ $x$ $\le$ 0.19), MnSi, ZrZn$_2$, Fe$_x$Co$_{1-x}$Si (0.36 $\le$ $x$ $\le$ 0.91), Ni$_x$Rh$_{1-x}$ (0.72 $\le$ $x$ $\le$ 0.90),  Pt$_{1-x}$Ni$_x$ (0.429 $\le$ $x$ $\le$ 0.502), Y$_2$Ni$_x$ (6.7 $\le$ $x$ $\le$ 7.0), YNi$_{2.9}$, YNi$_3$, Y$_2$Ni$_{17}$, Y$_2$Ni$_{15}$, Fe and Ni\cite{takahashi1,takahashi3,yang1,waki1}. 
 
 The data points for the actinide ferromagnets in Fig. 3 are distributed overwider parameter ranges of ${T_{\rm C}}/{T_0}$ and ${p_{\rm eff}}/p_{\rm s}$. This suggests that the $5f$ electrons in the actinide ferromagnets show the various degree of itinerancy. The data for the actinide ferromagnets are plotted close to those of the itinerant ferromagnets in the $3d$ transition metals and their intermetallics. The relations between the two quantities ${T_{\rm C}}/{T_0}$ and ${p_{\rm eff}}/p_{\rm s}$ in the actinide ferromagnets follow the relation in the Takahashi's spin fluctuation theory for $T_{\rm C}/T_0$ $<$ 1.0. This suggests itinerant character of the $5f$ electrons in most of the actinide ferromagnets. Similarities of the ferromagnetic properties between the $3d$ and $5f$ electrons systems are evident and suggest that the spin fluctuation theory can be applied to the $5f$ actinide ferromagnets. This finding is surprising since there are differences in the nature of the $3d$ and $5f$ electrons as mentioned in the Introduction. 
  
 Around ${T_{\rm C}}/{T_0}{\,}{\sim}{\,}1$, the data points of U$_3$TiSb$_5$, UCuP$_2$, UCuAs$_2$, and URh$_6$Ge$_4$ deviate from the theoretical relation. The situation $T_{\rm C}/{T_0}{\,}={\,}1$ corresponds to the local moment system in spin fluctuation theory and the deviation of the data points may be due to characteristic features of the localized $5f$ electrons not included in the theory as will be discussed in the next section. 

 We look for other systematic tendencies among the basic and the spin fluctuation parameters. Figure 4 shows relation between ${T_{\rm C}}/{T_0}$ and the spontaneous magnetic moment $p_{\rm s}$ and for the uranium, neptunium, and plutonium ferromagnets. The relation in the $3d$ system is also plotted. There is a general tendency of a positive correlation between the two quantities. The data points of the $3d$ systems are comparably scattered but there seems to be a linear relation between $p_{\rm s}$ and ${T_{\rm C}}/{T_0}$ in the uranium and neptunium ferromagnets. It is reasonable that $p_{\rm s}$ increases with increasing degree of localization of the $5f$ electrons. The value of  $p_{\rm s}$ in the uranium and neptunium ferromagnets increases as a function of ${T_{\rm C}}/{T_0}$ followed by a kink structure in the relation between the two quantities at $(T_{\rm C}/{T_0})_{\rm {kink}}{\,}{=}{\,}0.32{\,}{\pm}{\,}0.02$ as denoted by an arrow in the figure. The bold dotted line is a fit to the data of the uranium and neptunium ferromagnets for $T_{\rm C}/{T_0}{\,}{<}{\,}(T_{\rm C}/{T_0})_{\rm {kink}}$ with the function $p({T_{\rm C}}/{T_0})=a({T_{\rm C}}/{T_0})^{-n}$ where $a$ and $n$ are fitting parameters. The values of $a$ and $n$ are determined as 3.05 and 0.838, respectively. $p_{\rm s}$ increases more weakly above $(T_{\rm C}/{T_0})_{\rm {kink}}$. The data points are scattered around ${T_{\rm C}}/{T_0}{\,}{\sim}{\,}1$, similar to the ones between ${T_{\rm C}}/{T_0}$ and ${p_{\rm eff}}/p_{\rm s}$ shown in Fig. 3. It is interesting to note that the extrapolated value of $p_{\rm s}$ for ${T_{\rm C}}/{T_0}{\,}{\rightarrow}{\,}1$ with the fitted line is close to the values of effective Bohr magneton number of the free actinide ions (3.58 ${\mu}_{\rm B}$ for U$^{4+}$: $f^2$ configuration, 3.62 ${\mu}_{\rm B}$ for U$^{3+}$ and Np$^{4+}$: $f^3$ configuration, 2.83 ${\mu}_{\rm B}$ for Np$^{3+}$ and Pu$^{4+}$: $f^4$ configuration). This is naturally expected since the value of $p_{\rm s}$ in the local moment system (${T_{\rm C}}/{T_0}{\,}{=}{\,}1$) should be equal to the free actinide ion ones. Meanwhile, actual data points are smaller than the extrapolated line above $(T_{\rm C}/{T_0})_{\rm {kink}}$, suggesting a spin-compensating mechanism that will be discussed in the next section. From the four data points of the plutonium ferromagnets, the value of $p_{\rm s}$ seems to approach to that (0.84 ${\mu}_{\rm B}$) of the trivalent Pu ion (Pu$^{3+}$: $f^5$ configuration). 

 \subsection{Some technical notes}  
  We explain the analyses of several compounds. The analyses of most of the ferromagnets have been done using the magnetic data taken on single crystalline samples in magnetic field applied along the magnetic easy axes as mentioned in the previous section. This is because the ferromagnetic state in actinide system generally has the magnetic anisotropy. But there are some exceptions. The analysis has been done on magnetic data taken on polycrystalline samples in the literatures for UCo$_{1-x}$Ru$_x$Ge ($x$ = 0.10 and 0.23), UCo$_{1-x}$Fe$_x$Ge, Np$_2$C$_3$, and PuGa$_3$ marked with a dagger$^{\dagger}$ in Table II and III. The series of UCo$_{1-x}$Ru$_x$Ge, UCo$_{1-x}$Fe$_x$Ge, URh$_{1-x}$Co$_x$Ge, URh$_{1-x}$Ir$_x$Ge, and the ferromagnetic superconductors URhGe and UCoGe crystalize in the orthorhombic TiNiSi-type structure. The strong uniaxial anisotropy is a characteristic feature in the ferromagnetic state of this system. The spin fluctuations parameters of UCo$_{1-x}$Ru$_x$Ge ($x$ = 0.10 and 0.23) were obtained from the analyses of the magnetization data taken on polycrystalline samples multiplied by three. Note that the parameters for $x$ = 0.03, and 0.12 in UCo$_{1-x}$Ru$_x$Ge were obtained with the data taken on single crystalline samples. This method may be reasonable for this system considering the systematic changes of the basic and spin fluctuation parameters as a function of the concentration $x$ in the series. The same treatment has been done in the analysis for UCo$_{1-x}$Fe$_x$Ge\cite{huang1}. There has been no report for the effective paramagnetic moment $p_{\rm eff}$ in UCo$_{1-x}$Fe$_x$Ge. In this study, the value of $p_{\rm eff}$ is assumed to be 1.90. The effect of the alloying does not largely affect the value of $p_{\rm eff}$ in the doped systems of UCoGe or URhGe. The values of $p_{\rm eff}$ are between 1.83 and 1.93 ${{\mu}_{\rm B}}$/U in UCo$_{1-x}$Ru$_x$Ge, UCo$_{1-x}$Rh$_x$Ge, and UCoGe. This approximation of $p_{\rm eff}$ is reasonable but it could cause small uncertainty in $p_{\rm eff}/{p_s}$ that is reflected in error bars of the data points.

  The crystal structure of Np$_2$C$_3$ is cubic\cite{lam}. Thus, the magnetic data of Np$_2$C$_3$ measured using polycrystalline sample may not largely differ from that of a single crystalline sample. We analyzed the magnet without a correction. There are two types of crystal structure for PuGa$_3$. One is trigonal type ($R$-3$m$) and the other is hexagonal DO19 type ($P6_{3}/mmc$)\cite{boulet}. The former shows the ferromagnetic transition at $T_{\rm C}$ = 20 K and the latter the antiferromagnetic one at Neel temperature $T_{\rm N}$ = 24 K. The analysis was on the data taken on polycrystalline sample for the former trigonal PuGa$_3$. We estimate possible errors of the parameter shown in Figs. 3 and 4. considering several factors that give uncertainty in the analyses on data taken on polycrystalline samples.

\section{DISCUSSIONS}
 We have analyzed the basic magnetic data of the actinide ferromagnets with Takahashi's spin fluctuation theory and found that the theoretical relation in the theory applied to $3d$ system is satisfied also for most of the actinide ferromagnets except in several cases. This suggests that relevant factors for the magnetic properties in actinide ferromagnetism are related to the itinerancy of the $5f$ electrons. The deviation of some data points at $T_{\rm C}/{T_0}{\,}{\sim}{\,}1$ is expected since the theory assumes the local moment system at $T_{\rm C}/{T_0}{\,}{=}{\,}1$ where other factors arising from the localized character of the $5f$ electrons not included in the theory set in. In this section, we discuss the consequences of the present analysis from several points of views.

 First we discuss the applicability of spin fluctuation theory to the actinide $5f$ systems. To date, there has been no experimental study for the upper limit of ${T_{\rm C}/T_0}$ where the spin fluctuation theory can be applied. The result shown in Fig. 3 suggests that spin fluctuation theory is valid for $T_{\rm C}/T_0{\,}{<}{\,}1$. We check the applicability of the theory more carefully. Here, we discuss the relative change between the experimental data of $({p_{\rm eff}}/p_{\rm s})_{\rm exp.}$ and theoretical value from the relation ${p_{\rm eff}}/p_{\rm s}=1.4({T_{\rm C}}/{T_0})^{-2/3}$ [Eq. (11)] in Takahashi's spin fluctuation theory. The relative change is defined as ${\Delta}/({p_{\rm eff}}/p_{\rm s})_{\rm theory}$, where ${\Delta}$ = $({p_{\rm eff}}/p_{\rm s})_{\rm exp.}-f({T_{\rm C}}/{T_0})$ and $f({T_{\rm C}}/{T_0})=1.4({T_{\rm C}}/{T_0})^{-2/3}$. Figure 5 (a) shows the ${T_{\rm C}/T_0}$-dependence of ${\Delta}/f({T_{\rm C}}/{T_0})$. As can been seen from Fig. 5 (a), the mean value of ${\Delta}/f({T_{\rm C}}/{T_0})$ for ${T_{\rm C}/T_0}{\,}{<}{\,}1.0$ is negative, indicating a systematic tendency that the data points of $({p_{\rm eff}}/p_{\rm s})_{\rm exp.}$ are located lower than the theoretical expected values. To see the ${T_{\rm C}/T_0}$-dependence of ${p_{\rm eff}}/p_{\rm s}$ more clearly, we fit the data of ${p_{\rm eff}}/p_{\rm s}$ for ${T_{\rm C}/T_0}{\,}{\ll}{\,}1$ with the function $f'({T_{\rm C}}/{T_0})=a({T_{\rm C}}/{T_0})^{-2/3}$ where $a$ is a fitting parameter. The value of $a$ is determined as 1.23. Figure 5 (b) shows the ${T_{\rm C}/T_0}$-dependence of ${\Delta}{'}/f'({T_{\rm C}}/{T_0})$, where ${\Delta}'$ = $({p_{\rm eff}}/p_{\rm s})_{\rm exp.}-f'({T_{\rm C}}/{T_0})$. These results suggest that the relation $f({T_{\rm C}}/{T_0})=1.4({T_{\rm C}}/{T_0})^{-2/3}$ is satisfied for $T_{\rm C}/T_0{\,}{{\rightarrow}}{\,}1$. The $5f$ electrons in these ferromagnets should be basically regarded as being itinerant for $T_{\rm C}/{T_0}{\,}{<}{\,}1.0$.

 We discuss the coefficient ${[1/({10{C_{4/3}}{\rm d}y/{\rm d}t})]}^{-1/2}$ in the relation between ${p_{\rm eff}}/{p_{\rm s}}$ and $({T_{\rm C}}/{T_0})^{-2/3}$ expressed by Eq. (11). In the Takahashi's spin fluctuation theory, ${\rm d}y/{\rm d}t$ is weakly temperature dependent, which gives the Curie-Weiss like behavior of the magnetic susceptibility $\chi$. The coefficient weakly depends on magnetic properties specific to ferromagnets and temperature regions where the effective magnetic moment ${p_{\rm eff}}$ is determined from the temperature dependence of $\chi$. As mentioned in the previous section, the value of the coefficient is estimated as 1.4 from comparisons between the theory and experimental data on itinerant ferromagnets in the $3d$ systems\cite{takahashi3,takahashi4}. The results shown in Figs. 5 (a) and 5(b) suggest that the coefficient in actinide $5f$ systems is smaller than that in the $3d$ systems.

      \begin{figure}[t]
\includegraphics[width=8.5cm]{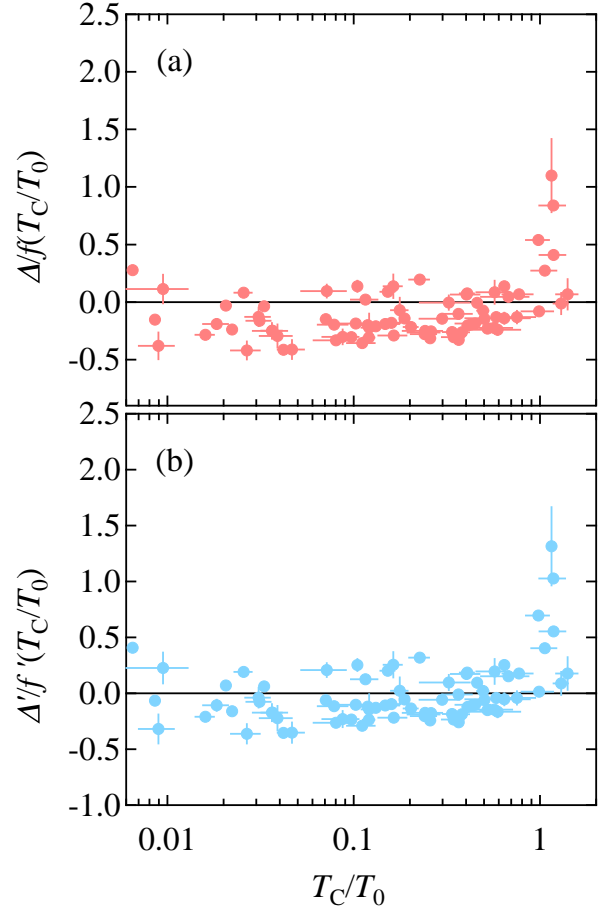}
\caption{\label{fig:epsart}(a)Relative change ${\Delta}/f({T_{\rm C}}/{T_0})$ as a function of ${T_{\rm C}/T_0}$ for the actinide ferromagnets. ${\Delta}= ({p_{\rm eff}}/p_{\rm s})_{\rm exp.}-f({T_{\rm C}}/{T_0})$ where $f({T_{\rm C}}/{T_0})=1.4({T_{\rm C}}/{T_0})^{-2/3}$ [Eq. (11)] from the Takahashi's spin fluctuation theory. (b)Relative change ${\Delta}{'}/f'({T_{\rm C}}/{T_0})$ as a function of ${T_{\rm C}/T_0}$ where ${\Delta}'$ = $({p_{\rm eff}}/p_{\rm s})_{\rm exp.}-f'({T_{\rm C}}/{T_0})$ and $f'({T_{\rm C}}/{T_0})=a({T_{\rm C}}/{T_0})^{-2/3}$. The value of $a$ is determined from the fit with the function $f'({T_{\rm C}}/{T_0})$ to the data points of $({p_{\rm eff}}/p_{\rm s})_{\rm exp.}$ for ${T_{\rm C}/T_0}{\,}{<}{\,}1.0$.}
\end{figure} 
      \begin{figure}[]
\includegraphics[width=8.3cm]{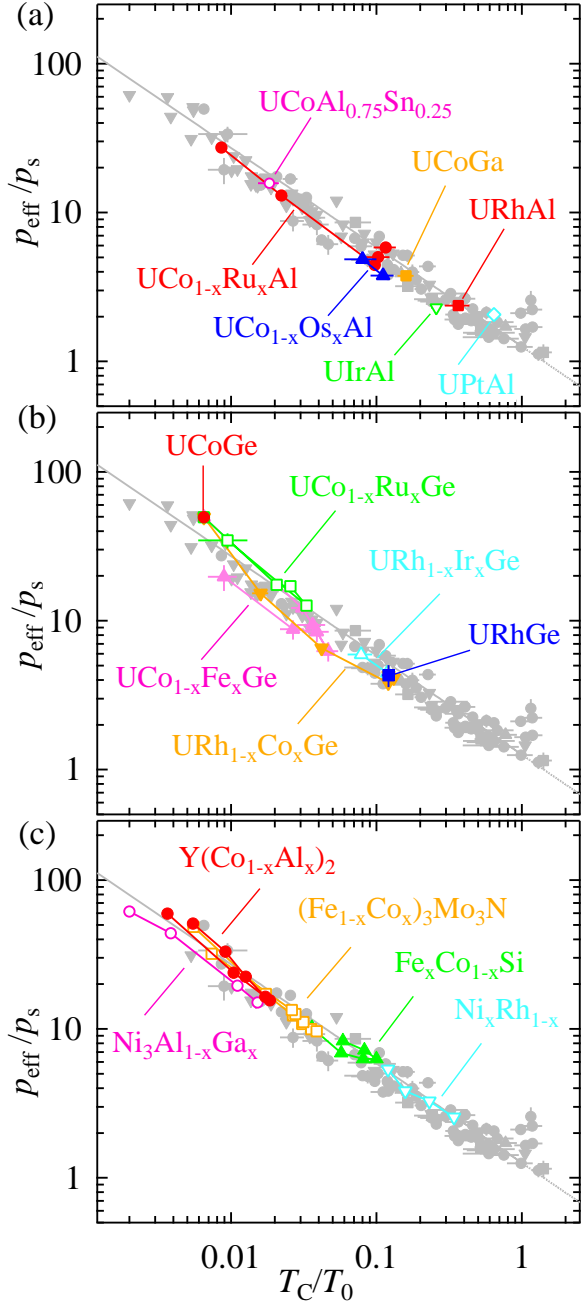}
\caption{\label{fig:epsart}
Relation between ${p_{\rm eff}}/p_{\rm s}$ and ${T_{\rm C}}/{T_0}$ in (a) a U{\it TX} ({\it T}: Co, Rh, Ir and Pt, {\it X}: Al, Ga, and Sn) series with hexagonal ZrNiAl-type crystal structure, (b) ferromagnetic superconductors URhGe and UCoGe and related doping systems with orthorhombic TiNiSi-type structure, and (c) itinerant ferromagents in the $3d$ system Y(Co$_{1-x}$Al$_x$)$_2$\cite{yoshimura1,yoshimura2}, Ni$_3$Al$_{1-x}$Ga$_x$\cite{yang1}, (Fe$_{1-x}$Co$_x$)$_3$Mo$_3$N\cite{waki1}, Fe$_x$Co$_{1-x}$Si\cite{shimizu}, and Ni$_x$Rh$_{1-x}$\cite{takahashi3}.
 }
\end{figure} 

 One important consequence of the present study is that Takahashi's spin fluctuation theory can be applied to the actinide ferromagnets whose spontaneous magnetic moment $p_{\rm s}$ is in the order of 1 ${\mu}_{\rm B}$/U. This may be reasonable because the theory assumes that the mean-square amplitude of the local spin fluctuation ${\langle}{S_{\rm L}^2}{\rangle}_{\rm {total}}$ is constant as a function of temperature. This is contrary to the early spin fluctuation theories whose applicability is limited to the weak coupling limit. The assumption here is of the constant ${\langle}{S_{\rm L}^2}{\rangle}_{\rm {total}}$ may be effective to the $f$-electron system of the rare-earth and the actinide compounds where the intra-atomic Coulomb interaction between the $f$ electrons is quite large. As mentioned in the introduction, Moriya's SCR theory was later extended to the $f$ electron system by the application of the constant ${\langle}{S_{\rm L}^2}{\rangle}_{\rm {total}}$ on the standard $s$-$f$ model\cite{moriya5}. The validity of the theory has been confirmed in a number of experimental studies\cite{lohneyseh,kambe1,kambe2,kambe3}.

 Figures 6 (a) and 6(b) show the relation between ${p_{\rm eff}}/p_{\rm s}$ and ${T_{\rm C}}/{T_0}$ in two uranium ferromagnetic systems: (a) U{\it TX} ({\it T}: Co, Rh, Ir and Pt, {\it X}: Al, Ga, and Sn) series with hexagonal ZrNiAl-type crystal structure and (b) ferromagnetic superconductors URhGe and UCoGe, and related doping systems with orthorhombic TiNiSi-type structure. The two systems have been extensively studied for the effect of alloying on the transition metal site that strongly affects the basic magnetic properties such as the spontaneous magnetic moment $p_s$ or the Curie temperature ${T_{\rm C}}$. A magnetic to nonmagnetic transition occurs at around $x$ = 0.77 in UCo$_{1-x}$Ru$_x$Al\cite{jirka1}, $x$ = 0.31 in UCo$_{1-x}$Ru$_x$Ge\cite{valivska1}, and $x$ = 0.22 in UCo$_{1-x}$Fe$_x$Ge\cite{huang1,huang2}, where anomalous physical properties around the ferromagnetic instability have been reported. It should be stressed here that the basic ferromagnetic properties are changed by the doping following the Takahashi's spin fluctuation theory similar to itinerant ferromagents in the $3d$ system Y(Co$_{1-x}$Al$_x$)$_2$\cite{yoshimura1,yoshimura2}, Ni$_3$Al$_{1-x}$Ga$_x$\cite{yang1}, (Fe$_{1-x}$Co$_x$)$_3$Mo$_3$N\cite{waki1}, Fe$_x$Co$_{1-x}$Si\cite{shimizu} and Ni$_x$Rh$_{1-x}$\cite{takahashi3} as shown in Fig. 6 (c). This is in contrast with the rare earth Ce or Yb $4f$ systems where competition between the magnetic inter-site RKKY interaction ${T_{\rm RKKY}}{\,}{\propto}{\,}{J{_{\rm cf}}^2}D({{\epsilon}_{\rm F}})$ and the demagnetizing Kondo effect ${T_{\rm K}}{\,}{\propto}{\,}{{\rm exp}[-1/J_{\rm cf}D({{\epsilon}_{\rm F}}})]$ has been described by the Doniach diagram\cite{doniach2}. Here, $J{_{\rm cf}}$ is the exchange constant between the $f$ and conduction electrons.

 Next, we discuss the deviations of the data points between $T_{\rm C}/T_0$ and ${p_{\rm eff}}/{p_{\rm s}}$ in U$_3$TiSb$_5$, UCuP$_2$, UCuAs$_2$, and URh$_6$Ge$_4$ with ${T_{\rm C}/{T_0}}{\,}{\sim}{\,}1.0$. The values of ${p_{\rm eff}}/p_{\rm s}$ of these ferromagnets are larger than unity. As mentioned before, spin fluctuation theory assumes localized magnetism at $T_{\rm C}/{T_0}{\,}={\,}1$ and the deviations may be due to the characteristic feature of the localized $5f$ electrons not included in the theory. We suggest a crystalline electric field effect (CEF) from surrounding ligand atoms on the $5f$ electrons can play an important role for the deviations as follows. The anisotropic magnetic property in UCu$_2$P$_2$, UCuP$_2$, and UCuAs$_2$ has been explained by the CEF effect on the localized $5f$ electrons\cite{kaczorowski1,kaczorowski5}. Let us discuss the effect of the CEF potential on the $5f$ electrons of the actinide ion based on the $LS$ coupling scheme\cite{santini1,santini2,moore,fazekas}. The CEF potential splits the degenerate $J$-multiplet of the actinide ions U$^{4+}$ ($5f^2$ configuration, total angular momentum $J$ = 4), U$^{3+}$ and Np$^{4+}$ ($5f^3$, $J$ = 9/2), Np$^{3+}$ and Pu$^{4+}$ ($5f^4$, $J$ = 4), and Pu$^{3+}$ ($5f^5$, $J$ = 5/2) into several separated energy levels (CEF level). The CEF effect depends on the site symmetry of the actinide ion in a crystal and surrounding ligand atoms. Generally, the value of the spontaneous magnetic moment $p_{\rm s}$ in the ferromagnetic state at $T$ = 0 K corresponds the expected value of $gJ_z$ of the CEF ground state and is smaller than that of the free actinide ion without the CEF effect. When the temperature is raised, the excited states of the CEF levels are populated and contribute to the magnetic property at finite temperature. Naturally, the effective magnetic moment $p_{\rm eff}$ at the higher temperature region is larger than $p_{\rm s}$. Of course, the ratio ${p_{\rm eff}}/p_{\rm s}$ depends on the CEF level scheme of each ferromagnet. This may be a reason of the deviations of the data points for $T_{\rm C}/{T_0}{\,}{\sim}{\,}1.0$. The scatterings of the data between ${T_{\rm C}}/{T_0}$ and $p_{\rm s}$ at around $T_{\rm C}/{T_0}{\,}{\sim}{\,}1.0$ shown in Fig. 4 may be also attributed to the CEF effect. There is no large deviation of the data from the theoretical relation for $T_{\rm C}/T_0{\,}{<}{\,}1$, suggesting that CEF effects on the $5f$ electrons becomes substantially weaker. There has been no report of observation of CEF excitation in the actinide ferromagnets for $T_{\rm C}/T_0{\,}{<}{\,}1$.
 
   The numbers of the actinide ferromagnets close to ${T_{\rm C}/{T_0}}{\,}{\sim}{\,}1.0$ is only about 10 \% of the total being analyzed in this study and the other ferromagnets follow the spin fluctuation theory. This fact suggests that the localized character of the $5f$ electrons is rare among the actinide ferromagents. It is noted that a CEF excitation in the inelastic neutron-scattering spectrum, evidence of the localized character of the $5f$ electrons, has been observed in only a few compounds such as UPd$_3$\cite{mcewen}, UPdSn\cite{nakotte2}, UGa$_2$\cite{kuroiwa}, and U$_3$Pd$_{20}$Si$_6$\cite{tateiwa2,tateiwa3, kuwahara}. It is interesting to consider the ``localized $5f$ electrons state" from experimental studies on UGa$_2$ and U$_3$Pd$_{20}$Si$_6$. The former is the ferromagnet analyzed in this study and the latter shows successive antiferromagnetic and ferromagnetic transitions at low temperatures\cite{tateiwa2}. Although the localized character of the $5f$ electrons has been suggested from the observations of the CEF excitations in the inelastic neutron-scattering studies, peak intensities are generally weak compared with rare-earth compounds\cite{kuroiwa,kuwahara}. The overall features of the nuclear spin-lattice relaxation rate 1/$T_1$ in NMR experiments cannot be interpreted based on the localized moment system for both compounds\cite{kambe4,maruta}, suggesting that the low-energy spin dynamics due to the hybridization between the $5f$ and conduction electrons. The nature of the localized state of the $5f$ electrons state may differ from that of the rare-earth $4f$ electrons systems. 
     
A number of actinide compounds show physical properties reminiscent of the Kondo effect as mentioned in the introduction. Among the actinide ferromagnets analyzed in this paper, the Kondo-like logarithmic temperature dependence of the resistivity (${\rho}{\,}{\sim}${\,}-ln$T$) has been reported in UCo$_{0.5}$Sb$_2$(${T_{\rm C}/T_0}$ = 0.240)\cite{tran1},  UPS(0.260)\cite{kaczorowski3}, UAuBi$_{2}$(0.407)\cite{rosa}, UCu$_{0.9}$Sb$_2$(0.448)\cite{bukowski1}, USbSe(0.676)\cite{kaczorowski6}, UTe(0.751)\cite{schoenes}, UCuAs$_{2}$(0.979)\cite{kaczorowski1,kaczorowski7}, USbTe(1.06)\cite{kaczorowski6}, UGa$_2$(1.12)\cite{honma}, UCuP$_2$(1.18)\cite{kaczorowski1}, NpNiSi$_2$(0.427)\cite{colineau2}, and NpFe$_4$P$_{12}$ (1.40)\cite{aoki1}. The coexistence of the Kondo effect and the ferromagnetic order in actinide systems has been theoretically studied with $S$ = 1 underscreened Anderson and Kondo lattice models\cite{coqblin1,coqblin2}. The minimum value of $T_{\rm C}/T_0$ among those of the ferromagnets is 0.240 for UCo$_{0.5}$Sb$_2$. It is interesting to note that this value is comparably close to $(T_{\rm C}/{T_0})_{\rm {kink}}$ (= 0.32 $\pm$ 0.02) where the kink is located in the relation between $T_{\rm C}/T_0$ and $p_{\rm s}$ as shown in Fig. 4. For $(T_{\rm C}/{T_0})_{\rm {kink}}{\,}{<}{\,}T_{\rm C}/{T_0}$, $p_{\rm s}$ increases more weakly as a function of $T_{\rm C}/T_0$ and the values of $p_{\rm s}$ are smaller than extrapolated ones with the fitted line in the uranium and neptunium ferromagnets. One possible interpretation is that $p_{\rm s}$ is suppressed by a Kondo-like spin-compensating mechanism. Some of the physical properties of the actinide compounds have been partly explained by taking into account Kondo or the CEF effect, but the two effects have not been well established theoretically in actinide metallic systems as mentioned in the Introduction. The main message from this analysis is that the $5f$ electrons in the actinide ferromagnets should be treated as being itinerant for $T_{\rm C}/{T_0}{\,}{<}{\,}1.0$. Further theoretical study is necessary to understand the dual nature of the $5f$ electrons in actinide systems.

 \section{Summary}
We have analyzed the magnetic data of 69 uranium, 7 neptunium and 4 plutonium ferromagnets with Takahashi's spin fluctuation theory. The analysis has been carried out using our experimental data for 17 uranium ferromagnets (group I). Data taken from the literature were analyzed for the remaining uranium (group II), neptunium, and plutonium ferromagnets. We have determined the basic and the spin fluctuation parameters of the ferromagnets and discuss the applicability of spin fluctuation theory to the actinide $5f$ electrons system. The ratio of the effective magnetic moment and the spontaneous one, ${p_{\rm eff}}/{p_{\rm s}}$, follows the generalized Rhodes-Wohlfarth relation, viz. ${p_{\rm eff}}/{p_{\rm s}}{\,}{\propto}{\,}({T_{\rm C}}/{T_0})^{-3/2}$ predicted by Takahashi's spin fluctuation theory in the actinide ferromagnets for $T_{\rm C}/{T_0}{\,}{<}{\,}1.0$, similarly to itinerant ferromagnets in the $3d$ transition metals and their intermetallics. This result suggests that the itinerant nature of the $5f$ electrons in the actinide ferromagnets and that the magnetic properties of the ferromagnets can be basically understood in the framework of the spin fluctuation theory. Meanwhile, data points between ${T_{\rm C}}/{T_0}$ and ${p_{\rm eff}}/{p_{\rm s}}$ deviate from the theoretical relation in several ferromagnets with ${T_{\rm C}/{T_0}}{\,}{\sim}{\,}1.0$, which may be due to the effect of the CEF on the $5f$ electrons. The value of the spontaneous magnetic moment $p_{\rm s}$ increases linearly as a function of ${T_{\rm C}}/{T_0}$ in the uranium and neptunium ferromagnets below $(T_{\rm C}/{T_0})_{\rm {kink}}{\,}{=}{\,}0.32{\,}{\pm}{\,}0.02$ where a kink structure appears in relation between the two quantities.  $p_{\rm s}$ increases more weakly above $(T_{\rm C}/{T_0})_{\rm {kink}}$. A possible interpretation with the ${T_{\rm C}}/{T_0}$-dependence of $p_{\rm s}$ is given in terms of the Kondo effect.

 \section{ACKNOWLEDGMENTS}
The authors thank Prof. A. V. Andreev and Dr. D. Gorbunov for providing their unpublished data of some uranium ferromagnets and Prof. Y. Takahashi for helpful advice on spin fluctuation theory. We acknowledge discussions with Dr. K. Kaneko, Dr. N. Metoki, Dr. S. Kambe, Dr. K. Deguchi, Prof. N. K. Sato, and Prof. H. Yamagami. We also thank Prof. Z. Fisk for discussions and his editing of this paper. This work was supported by Japan Society for the Promotion of Science (JSPS) KAKENHI Grant No. 16K05463, No. 16K05454, No. 16KK0106, Np. JP15H05884(J-Physics), and No. 26400341.

\newpage 
\bibliography{apssamp}

\end{document}